\documentclass[ste,twoside]{stefano}

\usepackage{amsmath}
\usepackage{amssymb}
\usepackage{graphics}
\usepackage{rotating}
\usepackage{cite}
\usepackage{color}

\def\makeheadbox{{%
\hbox to0pt{\vbox{\baselineskip=10dd\hrule\hbox
to\hsize{\vrule\kern3pt\vbox{\kern3pt \hbox{  {\sc Phys. Rev. A
{\bf 73}, 042107-7} } \hbox{ {\sc
{\color{blue}{dma}}[{\color{black}{imecc}}]{\color{red}{UniCamp}}
} \hspace*{10.4cm} {\color{blue}{$\boldsymbol{\Sigma \delta
\Lambda}$}} }
\kern3pt}\hfil\kern3pt\vrule}\hrule}%
\hss}}}

\def\0{\mbox{\tiny $0$}}
\def\1{\mbox{\tiny $1$}}
\def\2{\mbox{\tiny $2$}}
\def\3{\mbox{\tiny $3$}}
\def\4{\mbox{\tiny $4$}}
\def\5{\mbox{\tiny $5$}}
\def\6{\mbox{\tiny $6$}}
\def\7{\mbox{\tiny $7$}}
\def\8{\mbox{\tiny $8$}}
\def\9{\mbox{\tiny $9$}}
\def\a{\mbox{\tiny $(1)$}}

\def\c{\mbox{\tiny $(3)$}}

\def\R{\mbox{\tiny $R$}}

\def\B{\mbox{\tiny $B$}}

\def\N{\mbox{\tiny $N$}}
\def\I{\mbox{\tiny $I$}}
\def\II{\mbox{\tiny $II$}}
\def\III{\mbox{\tiny $III$}}
\def\infm{\mbox{\tiny $-\infty$}}
\def\infp{\mbox{\tiny $+\infty$}}
\def\m{\mbox{\tiny min}}

\begin{document}
%
%%%%%%%%%%%%%%%%%%%%%%%%%%%%%%%% PAPER %%%%%%%%%%%%%%%%%%%%%%%%%%%%%%%%%%%%%

\title{\large  BARRIER PARADOX IN THE KLEIN ZONE}
%\subtitle{}

\author{
%Alex E. Bernardini\inst{1}
Stefano De Leo\inst{1}
%\thanks{Partially supported by the FAPESP grant 99/09008--5.}
%\and
%Gisele C. Ducati\inst{1,2}
%\thanks{Supported by a CAPES PhD fellowship.}
%\and
%Celso C. Nishi\inst{2}
\and Pietro P. Rotelli\inst{2} }

\institute{
%Department of Cosmic Rays and Chronology, State
%University of Campinas\\
%PO Box 6165, SP 13083-970, Campinas, Brazil\\
%{\em alexeb@ifi.unicamp.br} \and
Department of Applied Mathematics, State University of Campinas\\
PO Box 6065, SP 13083-970, Campinas, Brazil\\
{\em deleo@ime.unicamp.br}
%{\em ducati@ime.unicamp.br}
%\and
%Department of Mathematics, University of Parana\\
%PO Box 19081, PR 81531-970, Curitiba, Brazil\\
%{\em ducati@mat.ufpr.br}
%\and Department of Cosmic Rays and Chronology, State University of
%Campinas\\
%PO Box 6165, SP 13083-970, Campinas, Brazil\\
%{\em ccnishi@ifi.unicamp.br}
\and
Department of Physics, INFN, University of Lecce\\
PO Box 193, 73100, Lecce, Italy\\
{\em rotelli@le.infn.it} }

%%%%%%%%%%%%%%%%%%%%%%%%%%%%%%%%%%%%%%%%%%%%%%%%%%%%%%%%%%%%%%%%%%%%%%%%%%%
%%%%%%%%%%%% DATE ABSTRACT PACS % %%%%%%%%%%%%%%%%%%%%%%%%%%%%%%%%%%%%%%%%%

\date{Submitted: {\em January, 2006} - Revised:  {\em March, 2006} }
% Warning: Where is the date?

\abstract{We study the solutions for a one-dimensional
electrostatic potential in the Dirac equation when the incoming
wave packet exhibits the Klein paradox (pair production). With a
barrier potential we demonstrate the existence of multiple
reflections (and transmissions). The antiparticle solutions which
are necessarily localized within the barrier region create new
pairs with each reflection at the potential walls.
 Consequently we encounter a new "paradox" for the barrier because
successive outgoing wave amplitudes grow geometrically.}

%%%%%%%%%%%%%%%%%%%%%%%%%%%%%%%%%%%%%%%%%%%%%%%%%%%%%%%%%%%%%%%%%%%%%%%
%%%%%%%%%%%%%%%%%%%%%%%%%%%%%%%%%%%%%%%%%%%%%%%%%%%%%%%%%%%%%%%%%%%%%%%

%%%%%%%%%%%%%%%%%%%%%%%%%%%%%%%%%%%%%%%%%%%%%%%%%%%%%%%%%%%%%%%%%%%%%%%
%%%%%%%%%%%%%%%%%%%%%%%%%%%%%%%%%%%%%%%%%%%%%%%%%%%%%%%%%%%%%%%%%%%%%%%

%\PACS{ {03.65.Xp}{}}

% Warning: No PACS code given

%02.10.Hh Rings and algebras
%02.10.Ud Linear algebra
%02.10.Yn Matrix theory

%02.30.Hq Ordinary differential equations
%02.30.Jr Partial differential equations
%02.30.Tb Operator theory

%03.65.-w Quantum mechanics
%03.65.Ca Formalism
%03.65.Ta Foundations of quantum mechanics; measurement theory
%03.65.Xp Tunnelling, traversal time,quantum Zeno dynamics
%03.65.Pm Relativistic wave equations

%12.15.F Quarks and lepton masses and mixing
%14.60.Pq Neutrino mass and mixing

%\offprints{~Stefano De Leo.}

\titlerunning{\sc barrier paradox in the klein zone}

\maketitle

%%%%%%%%%%%%%%%%%%%%%%%%%%%%%
%%%%%%%%%%%%%%%%%%%%%%%%%%%%%
\section*{I.  INTRODUCTION.}
%%%%%%%%%%%%%%%%%%%%%%%%%%%%%
%%%%%%%%%%%%%%%%%%%%%%%%%%%%%

In this paper, we consider two related one dimensional
(electrostatic) square potentials: the step and barrier potential
within the Dirac equation. We shall analyze them for the case when
the incoming energy (energies for wave packets) are in what we
shall call the \emph{Klein zone}, i.e. when $ E < V_{\0} - m$,
where $V_{\0}$ is the step/barrier height and $m$ the particle
mass. This is the situation in which only oscillatory solutions
exist throughout and where the so called Klein paradox reigns for
the step\cite{KLE29}.

They will be analyzed with use of the stationary plane wave
method\cite{SPM1,SPM2,COHEN}. However, when appropriate, we shall
also use wave packet arguments and terminology such as group
velocity, arrival times etc. For brevity, we will not recall here
the underlying formalism of convolution integrals or stationary
phase methods. We hope that these switches from time independent
to time dependent viewpoints, even if in the same sentence, will
not lead to any confusion. Actually, this is quit a common
practice and occurs for example whenever two contributing plane
wave solutions are referred to individually as "incoming" and/or
"outgoing".

We start with the step potential by recalling in the next Section
the arguments which lead to the Klein paradox,
 in which the reflection probability is higher
than the incoming probability. This is really no longer a paradox
since it is universally interpreted as due to the creation of a
particle-antiparticle pair at the potential
discontinuity\cite{HR81,SMG82,H98,NKM99,KSG04}. For an
electrostatic potential the antiparticle "sees" a potential dip
where the particle sees a potential rise and
vice-versa\cite{GMR85,GROSS}. This explains why oscillatory
solutions appear in the plane wave analysis both for free regions
(zero potential) and non. The below potential oscillatory
solutions are thus identified with \emph{physical} (above
potential) antiparticles. The antiparticles will be shown to have
energy $-E$ over a potential of $-V_{\0}$, so that with respect to
the potential free region they also lie in a Klein zone i.e., $-E
< 0 - m $ since necessarily the relativistic energy satisfied $ E
> m $ . Even the alternative candidate step solution in which the
reflection probability is less than the incoming probability can
only be understood by \emph{ pair annihilation} at the step.
However, this solution necessarily implies the existence of an
incoming antiparticle in addition to the incoming particle and
thus is rejected because it violates the initial conditions
assumed.

As for the barrier, a plane wave analysis yields only \emph{one}
solution, that with the sum of the reflected and transmitted
probabilities (both positive) lower than the incoming
probability\cite{TM91,NKM93}. Not only does this solution appear
inconsistent with the step result (Klein paradox) but it seems, to
us, not interpretable in terms of pair production and/or
annihilation. To try and understand the situation, we shall then
apply a procedure we have called the two step approach to the
barrier. We previously applied it to above barrier
diffusion\cite{MPLA}. In that case it exemplified the presence of
multiple wave packets\cite{A89}, due to multiple reflections,
which only in the limit of \emph{complete overlap} reproduced the
plane wave barrier result and consequent resonance effects.

While the two step analysis guaranties consistency between the
step and barrier because it uses the former for the calculation of
the latter, it will lead us to a new kind of paradox. Any
antiparticle created at either of the steps that form the barrier
will necessarily lie entrapped in the  barrier region (which it
sees as a well). It will bounce back and forth indefinitely. Since
it also satisfies the Klein condition, it will have a non zero
probability of creating antiparticle-particle pairs at each
reflection. This means that the density of antiparticles in the
well will grow at each reflection. A corresponding geometric
growth  of the multiple reflected and transmitted particle
probabilities also occurs.

In Appendix, we collect the relevant Dirac solutions and define
our conventions\cite{ZUB}.  This is done in three dimensions,
although for our calculations we consider only one dimensional
potentials along the z axis. We shall also assume that the
incoming  spin is polarized along the same axis (i.e. a positive
helicity eigenstate). Since all spin flip terms can readily be
shown to be absent, we shall neglect them and this will simplify
our continuity equations. The  Appendix  also contains  the Dirac
solutions in the presence of a non zero, but constant, potential
which we will indeed use.

In the next Section we describe the solutions to the step
potential and in particular the Klein paradox. These
demonstrations can be found in numerous
articles\cite{HR81,SMG82,H98,NKM99,KSG04} and
textbooks\cite{GMR85,GROSS,ZUB}, but with an important proviso.
The traditional approach uses "positive energy" (above-potential)
spinors under the step. We, on the contrary, employ the
below-potential solutions there. Nevertheless, the results for
{\em probabilities} turn out to be the same. In Section III, we
present our plane wave analysis of the barrier. We describe the
difficulties of the interpretation of this solution in terms of
pair creation/annihilation. Section IV repeats this analysis but
with the two step method. An important revaluation of the results
of the previous Section can then be made. We conclude in Section V
by a resume of our results and a discussion of the predicted
 unlimited growth of antiparticle and particle wave packets densities.

%%%%%%%%%%%%%%%%%%%%%%%%%%%%%%%%%%%%%%%%%%%%%%%%%%%%%%%%%
%%%%%%%%%%%%%%%%%%%%%%%%%%%%%%%%%%%%%%%%%%%%%%%%%%%%%%%%%
\section*{II. THE SINGLE STEP.}
%%%%%%%%%%%%%%%%%%%%%%%%%%%%%%%%%%%%%%%%%%%%%%%%%%%%%%%%%
%%%%%%%%%%%%%%%%%%%%%%%%%%%%%%%%%%%%%%%%%%%%%%%%%%%%%%%%%
Our potential model is one-dimensional with the $z$-axis chosen as
the privileged space direction. The potential is chosen as
\vspace*{.5cm}

\begin{picture}(180,90)
\thinlines \put(50,10){\vector(0,1){66}} \put(47,81){$V$}
\put(180,55){$V_{\0}$}
\put(180,30.8){$V_{\0}-2m$}\put(50,10){\vector(1,0){135}}
\put(187,8){$z$} \put(48,0){$0$} \put(-2,70){\mbox{\small \sc
Partial}} \put(-2,62){\mbox{\small \sc Reflection}}
\put(-2,47){\mbox{\small \sc Total}} \put(-2,39){\mbox{\small \sc
Reflection}} \put(-2,23){\mbox{\small \sc Klein}}
\put(-2,15){\mbox{\small \sc Zone}}
 \thicklines
 \put(2,10){\line(1,0){48.4}}
\put(50,10){\line(0,1){48.5}} \put(0,55){$-----$}
 \put(0,30.8){$-----$}
 \put(50,58){\line(1,0){128}}
 \put(108,65){$E>V_{\0}+m$}
\put(63,42){$V_{\0}-m<E<V_{\0}+m$} \put(85,19){$m<E<V_{\0}-m$}
 \put(50,30.8){$-------------$}
 \put(245,42){$
V(z) = \left\{
\begin{array}{rll}
0\,\,, &  \,\,\,z < 0\,\,, &\,\,\,\mbox{\small \sc region I}\,\,, \\
V_{\0}\,\,, &\,\,\,z>0\,\,,&\,\,\,\mbox{\small \sc region II}\,\,.
\end{array} \right.
$}
\end{picture}

\vspace*{.5cm}

 \noindent For the purpose of this paper, the energy of each plane wave
 lies by assumption in the Klein zone $ E < V_{\0} - m$. We also assume that
 the incoming wave is a positive helicity
 positive energy solution. Ignoring, for simplicity, the spin flip terms
which can easily be shown not to exist\cite{ZUB}, the reflected
wave is consequently a negative helicity (because of the change in
direction) outgoing particle. The solutions in the free zone can
thus be written as,
\begin{equation}
\Psi_{\I}(z,t)= \left\{\, u^{\a}(p\,;0)\,\, \exp[\,i\, p\,z\,] +
R\, u^{\a}(-p\,;0)\,\,
\exp[\,-\,i\,p\,z\,]\,\right\}\,\exp[\,-\,i\,E\,t\,]\,\,,
\end{equation}
with
\[u^{\a}(p\,;0)= \left(\,\begin{array}{c}1\\0\\ \displaystyle{\frac{p}{E+m}}\\0
\end{array}\,\right)\,\,,\]
and $p=\sqrt{E^{^{\,\2}} - m^{\2}}$. By choice we use unnormalized
spinors  which is equivalent to the convention of absorbing any
normalization factors into the coefficients $1$, $R$, and $T$. The
reflection coefficient is $R$. The reflection probability is
consequently $|R|^{^{\2}}$. It should be warned that some other
authors use instead the letter $R$ for the reflection probability.

The solution \emph{under} the step (see Appendix) can initially be
taken as either of two forms differing in the sign of the
momentum. Not \emph{both} because we expect there to be only one
outgoing wave in the region with $z
> 0 $ in analogy with the above step case. We shall consider
them one at a time since it is by no means obvious \emph{which} of
them represents a physical outgoing (right moving) object.
Consider first the $ u^{\c}(q\,;V_{\0})\,\exp[\,i\,q\,z]$
solution, with transmission coefficient T,
\begin{equation}
\Psi_{\II}(z,t)= T\,u^{\c}(q\,;V_{\0})\,\, \exp[\,i\,
q\,z\,]\,\exp[\,-\,i\,E\,t\,]\,\,,
\end{equation}
where
\[u^{\c}(q\,;V_{\0})= \left(\,\begin{array}{c}
-\,\displaystyle{\frac{q}{|E-V_{\0}|+m}}\\0\\1\\0
\end{array}\,\right)\,\,,\]
and $q=\sqrt{(E-V_{\0})^{^{\2}} - m^{\2}}$.

Continuity at $ z=0 $ yields:
\begin{eqnarray*}
 1+R & = & -\, \dfrac{q}{|E-V_{\0}|+m}\,\,T\,\,, \\
 1-R & = & \dfrac{E+m}{p}\,\,T\,\,.
 \end{eqnarray*}
 We note that with our choice of spinors we have just two equations
in two unknowns and the fact that a solution to these equations
exist is confirmation, a posteriori, of the absence of spin flip.
The solution is
\begin{equation}
R=\frac{\alpha+1}{\alpha-1}\hspace*{.5cm}\mbox{and}\hspace*{.5cm}
T=\sqrt{\frac{E-m}{E+m}}\, \,\frac{2}{1-\alpha}\,\,,\end{equation}
   with $\alpha=\sqrt{(|E-V_{\0}|-m)(E-m)}\,/
\sqrt{(|E-V_{\0}|+m)(E+m)}$.

 Since we are in the Klein zone
$|E-V_{\0}| > m $ and hence $ 1 > \alpha > 0 $. This means that $
R < -1 $ and that the reflection probability is $|R|^{^2} > 1$,
i.e. exceeds the incoming probability (Klein paradox). The direct
calculation of the transmitted probability is not $|T|^{^2}$
because we are in a region with a different value of potential to
the incoming wave. The transmitted probability is given by
\begin{eqnarray}
\frac{q}{|E-V_{\0}|}\,\left|\,T\,u^{\c}(q\,;V_{\0})\,\right|^{\2}\,\mbox{\Large
$/$}\,\frac{p}{E}\,\left|\,u^{\a}(p\,;0)\,\right|^{\2} &= &
\frac{q}{|E-V_{\0}|}\,\frac{E-m}{E+m}\,\frac{4}{(1-\alpha)^{^{\2}}}\,
\frac{2\,|E-V_{\0}|}{|E-V_{\0}|+m}\,\frac{E}{p}\,\frac{E+m}{2\,E}\nonumber
\\
 & = & \frac{4\,\alpha}{(1-\alpha)^{^{\2}}}\nonumber \\
  & = & |R|^{^2} -1\,\,.
\end{eqnarray}
The same result is obtained by using the traditional spinor
$u^{\a}$ under the step\cite{ZUB}. This latter choice of spinor is
formally incorrect and has the unpleasant feature of containing a
denominator $(E-V_{\0})+m $ which tends to zero in the
(antiparticle) rest frame limit (limit of the Klein zone) when
$[-E - (- V_{\0})]\to m$. However, it is just this vanishing
denominator which inverts the "small component" with the "large
component" in the non-relativistic situation, and consequently
yield the same results as that above.

Now, let us interpret the Klein paradox in \emph{physical} terms.
Since the incoming particle is assumed charged (recall that the
potential is of an electrostatic nature) this means that the
reflected wave packet carries more charge than the incoming one.
Extra particle charge has been created at reflection. Since charge
is a conserved quantum number the wave packet in the positive $z$
region must be of opposite charge i.e., represent an antiparticle
wave packet travelling above a potential trough. It is a basic
axiom of our interpretation that the below barrier solutions {\em
cannot} be particles, they must represent antiparticles or
"holes".  Hence, \emph{pair creation} has occurred, or more
precisely has a probability of occurring. What is more, the
antiparticle wave packet \emph{must move to the right }since it
must exist for all future times to guarantee charge
conservation\cite{HR81,H98,SMG82,NKM99,KSG04}.

The following picture shows the antiparticle viewpoint. The
potential flips sign because the antiparticle's charge is opposite
to that of the particle. The energy of the antiparticle is
determined as follows:

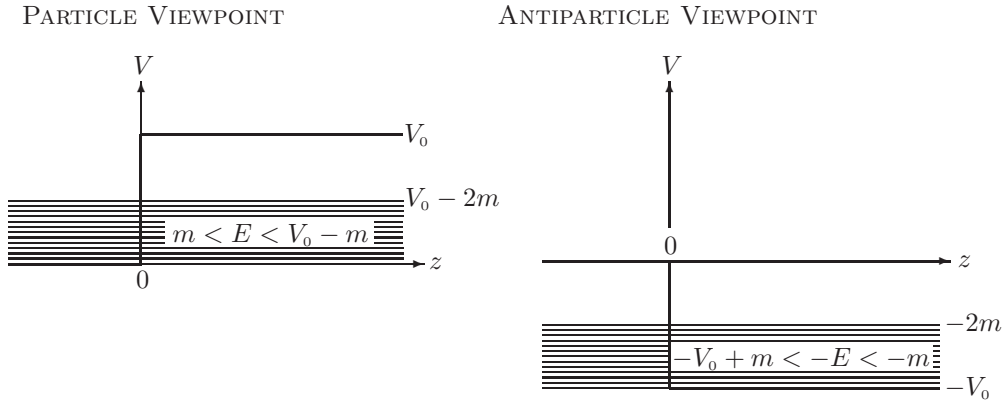
\vspace*{1cm}
\begin{picture}(380,100) \thinlines
\put(50,10){\vector(0,1){68}} \put(47,81){$V$}
\put(149.5,55){$V_{\0}$} \put(150,31){$V_{\0}-2m$}
\put(62,17.5){$m<E<V_{\0}-m$} \put(0,9){\line(1,0){68}}
\put(50,9){\vector(1,0){107}} \put(159,7){$z$} \put(48.25,0){$0$}
\put(0,33){\line(1,0){68}} \put(123.5,33){\line(1,0){25.5}}
 \thicklines
 \put(0,9){\line(1,0){50}}\put(50,9){\line(0,1){49}}
\put(50,58){\line(1,0){99}} \thinlines
 \put(0,31){\line(1,0){149}}
\put(0,33){\line(1,0){149}}
 \put(0,29){\line(1,0){149}}
 \put(0,27){\line(1,0){149}}
 \put(0,15){\line(1,0){149}}
 \put(0,13){\line(1,0){149}}
 \put(0,11){\line(1,0){149}}
\put(0,25){\line(1,0){59}} \put(0,23){\line(1,0){59}}
\put(0,21){\line(1,0){59}} \put(0,19){\line(1,0){59}}
\put(0,17){\line(1,0){59}} \put(138.5,25){\line(1,0){10.5}}
 \put(138.5,23){\line(1,0){10.5}}
  \put(138.5,21){\line(1,0){10.5}}
   \put(138.5,19){\line(1,0){10.5}}
    \put(138.5,17){\line(1,0){10.5}}
 \put(250,23){\vector(0,1){55}} \put(247,81){$V$}
\put(354,-40.5){$-V_{\0}$} \put(354,-16.5){$-2m$}
\put(202,10){\vector(1,0){154}} \put(359,8){$z$}
\put(248.,13){$0$} \thicklines \put(202,10){\line(1,0){48.5}}
\put(250,10){\line(0,-1){48}} \put(250,-38){\line(1,0){102}}
\put(251,-29.5){$-V_{\0}+m<-E<-m$}
 \thinlines
 \put(202,-38){\line(1,0){150}}
 \put(202,-32){\line(1,0){150}}
\put(202,-36){\line(1,0){150}} \put(202,-34){\line(1,0){150}}
\put(202,-14){\line(1,0){150}} \put(202,-16){\line(1,0){150}}
\put(202,-18){\line(1,0){150}} \put(202,-20){\line(1,0){150}}
\put(202,-22){\line(1,0){48}} \put(202,-24){\line(1,0){48}}
\put(202,-26){\line(1,0){48}} \put(202,-28){\line(1,0){48}}
\put(202,-30){\line(1,0){48}} \put(350,-22){\line(1,0){2}}
\put(350,-24){\line(1,0){2}} \put(350,-26){\line(1,0){2}}
\put(350,-28){\line(1,0){2}} \put(350,-30){\line(1,0){2}}
\put(5,100){\sc Particle Viewpoint} \put(185,100){\sc Antiparticle
Viewpoint}
\end{picture}

\vspace*{2cm}

\noindent The energy of the "particle" represented by $ u^{\c}$
is, by fiat, that of the incoming particle i.e., $E$. The
"particle" lies under the potential $V_{\0}$.
 The modulus of its momentum, is the
expression for $q$ given above.  After reinterpretation as a
physical antiparticle the magnitude of its momentum  must still be
given by $q$. Since the antiparticle sees a potential of $ -V_{\0}
$ its energy, say $E_a$ must satisfy $[E_a-(-V_{\0})]^{^{\2}} = (E
- V_{\0})^{^{\2}}$. From which we conclude therefore that $E_a=-E$
(see the above figures). This means that \emph{energy has been
conserved in the pair creation}. Because this has occurred at a
potential boundary there is no threshold energy needed. A similar
interpretation occurs with the Klein-Gordon equation\cite{GROSS}.
However, there is a difficulty  with the above interpretation. A
direct calculation of the group velocity of the below potential
"particle" solution yields,
\[
\frac{\mbox{d}z}{\mbox{d}t}
=\frac{\mbox{d}E}{\mbox{d}q}=\frac{q}{E-V_{\0}}\,\,.
\]
This is \emph{negative} because $E<V_{\0}$. We can only reconcile
this result with a right-moving antiparticle by invoking the
Feynmann-Stuckelburg rule that below-barrier solutions travel
backward in time. Whence $\mbox{d}t < 0 $ and consequently
$\mbox{d}z > 0$. The corresponding antiparticle wave function has
$\mbox{d}t > 0 $ as have all physical particles and hence must
exhibit a positive group velocity.

We take this opportunity to observe that the conventional charge
conjugate solution  differs in its space-time structure from that
of the "particle" by an overall complex conjugation. This has
\emph{no effect} upon the calculation of the group velocity. Thus
the charge conjugate wave function cannot represent correctly the
antiparticle state (it would yield the wrong sign for the group
velocity)\cite{SAK}. Another important observation, even if
obvious, is that to satisfy the continuity equations we need at
least three "touching" plane waves. Pair creation in the absence
of an incoming wave is not a solution.

Finally we consider, briefly, the alternative plane wave solution
under the step:
\begin{equation}
\Psi'_{\II}(z,t)= T'\,u^{\c}(-q\,;V_{\0})\,\, \exp[-\,i\,
q\,z\,]\,\exp[\,-\,i\,E\,t\,]\,\,.
\end{equation}
This yields the following solution to the continuity equations,
\begin{equation}
R'=\frac{\alpha-1}{\alpha+1}\hspace*{.5cm}\mbox{and}\hspace*{.5cm}
T'=\sqrt{\frac{E-m}{E+m}}\, \,\frac{2}{\alpha +1}\,\,.
\end{equation}
with $\alpha$ given as above.  Now $|R'| < 1$ and part of the
incoming charge has been \emph{annihilated}. Thus this
below-barrier solution \emph{must} involve an incoming
antiparticle from the right. It is therefore inconsistent with the
assumed initial boundary conditions and is consequently rejected.
It is the Klein paradox solution which is generally considered
valid for the step.

%%%%%%%%%%%%%%%%%%%%%%%%%%%%%%%%%%%%%%%%%%%%%%%%%%%%%%%%%
%%%%%%%%%%%%%%%%%%%%%%%%%%%%%%%%%%%%%%%%%%%%%%%%%%%%%%%%%
\section*{III THE PLANE WAVE BARRIER ANALYSIS.}
%%%%%%%%%%%%%%%%%%%%%%%%%%%%%%%%%%%%%%%%%%%%%%%%%%%%%%%%%
%%%%%%%%%%%%%%%%%%%%%%%%%%%%%%%%%%%%%%%%%%%%%%%%%%%%%%%%%

For the barrier,
\[
V(z) = \left\{
\begin{array}{rrl}
0\,\,, &   z < 0\,\,, &\,\,\,\,\, \mbox{\small \sc region I}\,\,, \\
V_{\0}\,\,, & \,\,\,\,0<z<l\,\,, &\,\,\,\,\, \mbox{\small \sc region II}\,\,,  \\
0\,\,, &   \,\,\,z > l\,\,, & \,\,\,\,\, \mbox{\small \sc region
III}\,\,.
\end{array} \right.
\]
We now derive the plane wave solution. In the left free region the
incoming and reflected solutions yield the combined wave function,
\begin{equation}
\Psi_{\I}(z,t)= \left\{\, u^{\a}(p\,;0)\,\, \exp[\,i\, p\,z\,] +
R_{\B}\, u^{\a}(-p\,;0)\,\,
\exp[\,-\,i\,p\,z\,]\,\right\}\,\exp[\,-\,i\,E\,t\,]\,\,,
\end{equation}
with $R_{\B}$ the barrier reflection coefficient. The solutions
"under" the potential give
\begin{equation}
\Psi_{\II}(z,t)= \left\{\, A_{\B}\,u^{\c}(q\,;V_{\0})\,\,
\exp[\,i\, q\,z\,] + B_{\B}\,u^{\c}(-q\,;V_{\0})\,\, \exp[-\,i\,
q\,z\,]\,\right\}\,\exp[\,-\,i\,E\,t\,]\,\,.
\end{equation}
Such solutions represent right ($A_{\B}$) and left ($B_{\B}$)
moving  antiparticles. The latter are now allowed as a consequence
of reflection at the second potential discontinuity. The solution
in the free region beyond the barrier is given by a single
outgoing plane wave,
\begin{equation}
\Psi_{\III}(z,t)= T_{\B}\, u^{\a}(p\,;0)\,\exp[\,i\,
p\,z\,]\,\exp[\,-\,i\,E\,t\,]\,\,,
\end{equation}
with $T_{\B}$ the transmission coefficient.

 The continuity
conditions, best displayed in matrix form are thus,
\begin{eqnarray*}
\left( \begin{array}{rr} 1 & 1 \\ 1 & \,\,-1 \end{array} \right)\,
\left( \begin{array}{c} 1 \\R_{\B} \end{array} \right) & = &
\sqrt{\frac{E+m}{E-m}}\,\left( \begin{array}{rr} -\alpha & \,\,\,\,\alpha \\
1 &  1 \end{array} \right)\,\left( \begin{array}{c} A_{\B}
\\B_{\B}
\end{array} \right)\\
\left( \begin{array}{rr} -e^{iq\,l}  & \,\,\,\,e^{-iq\,l} \\
e^{iq\,l}& e^{-iq\,l}  \end{array} \right)\,\left(
\begin{array}{c} A_{\B}
\\B_{\B}
\end{array} \right) & = & \sqrt{\frac{|E-V_{\0}|+m}{|E-V_{\0}|-m}}\,
\left( \begin{array}{c} 1\\ \alpha
\end{array} \right)\,T_{\B} \,e^{ipl}
\end{eqnarray*}
where $\alpha$ is as defined in the previous Section. The
solutions of these equations is straightforward,
\begin{equation}
R_{\B} =i\,\frac{\alpha^{\2} -1}{2\,\alpha}\, \sin(ql)\,T_{\B}
\,e^{ip\,l}\hspace*{.5cm}\mbox{and}\hspace*{.5cm} T_{\B} =
e^{-ip\,l}\,\mbox{\large $/$}\,  \left[\,\cos(ql) + i\,
\frac{\alpha^{\2}+1}{2\,\alpha}\,\sin(ql)\,\right]\,\,.
\end{equation}
 Consequently, both
reflected and transmitted probabilities are positive and less than
 one (the incoming probability). Probability is conserved,
\begin{equation}
|R_{\B}|^{^{\2}} + |T_{\B}|^{^{\2}} = 1\,\,.
\end{equation}

At first sight this solution seems perfectly acceptable and even
conventional as long as one does not try to interpret it in
physical terms i.e., as long as one does not look into region II.
Notice that it is characterized by resonance phenomena when the
sine term in the denominator of $T_{\B}$ vanishes, in which cases
the transmission probability becomes unity. However, how can this
solution be compatible with the step analysis, and specifically
with the Klein paradox. If the barrier is extended indefinitely we
should in some way tend to the step solution and pair production.
This is what occurs for above barrier diffusion, albeit in a non
trivial manner. In the above barrier case,\emph{ multiple
reflections} occur for the barrier whereas only a single
reflection occurs for the step potential\cite{MPLA}. It is the
\emph{first} reflected wave packet of the barrier that coincides
with the step result (including its instantaneous reflection). One
could say that the subsequent wave packets for the barrier exist
but  are delayed indefinitely as the barrier length grows without
limit. A long barrier (with respect to the incoming packet width)
thus reproduces the step diffusion result for a finite time after
impact of the incoming particle.

Let us try to interpret the results of this Section. Since the
reflection probability is less than the incoming probability
particle charge in region I has decreased. Pair annihilation has
occurred at the origin discontinuity. The annihilating
antiparticle can only have come from pair creation at the second
discontinuity at a time previous to its annihilation. Therefore
the outgoing particle in region III had also to be created at this
earlier time. However, here we already have a problem since the
creation of a pair without an incoming contribution is
incompatible with the step continuity equations as already noted
in the previous Section. Is the (first) antiparticle produced at
the first or second discontinuity? If it where at the first,
coincident with the incoming particle we would have no problem
with continuity but we would have a problem with charge
conservation since we are not in accordance with the Klein
paradox. If at the second discontinuity we face the
afore-mentioned violation of continuity since there is no incoming
particle in region III.

 In the following Section we
shall approach the barrier as a two step process and consequently
reinterpret the above results.

%However, one could argue that one should reject the above barrier
%result because of a basic flaw. The plane wave analysis requires
%for its validity that all time phases are the same in each region
%and can hence be factored out. It is a\emph{ stationary analysis}.
%However, the below potential solution is anomalous because of the
%direction of time flow, or from a different view point because the
%(physical) antiparticle has a different energy $ -E $ from that of
%the particles and hence again a different time phase. From either
%viewpoint the time dependence in region II is different from that
%of the free regions. A collaborating argument is that to have the
%correct (finite) antiparticle group velocity in the well(step)
%potential the space-time phase factor must be given by
%$\exp[\,i\,(\,q\,z+E\,t\,)]=\exp[\,i\,(\,q\,z-E_a\,t)]$.

%%%%%%%%%%%%%%%%%%%%%%%%%%%%%%%%%%%%%%%%%%%%%%%%%%%%%%%%%
%%%%%%%%%%%%%%%%%%%%%%%%%%%%%%%%%%%%%%%%%%%%%%%%%%%%%%%%%
\section*{IV. THE TWO STEP APPROACH.}
%%%%%%%%%%%%%%%%%%%%%%%%%%%%%%%%%%%%%%%%%%%%%%%%%%%%%%%%%
%%%%%%%%%%%%%%%%%%%%%%%%%%%%%%%%%%%%%%%%%%%%%%%%%%%%%%%%%

The two step approach to the barrier actually involves the
calculation and use of three step potentials. One is that already
calculated in Section II at $z = 0$, for a left incoming particle,
characterized by $|R|>1$. Another is for a left incoming
antiparticle reflected from the potential discontinuity at $z=l$.
The third is for the reflected antiparticle impinging upon the
first potential discontinuity at the origin and hence arriving
from the right. Since these antiparticles are themselves in the
Klein zone for the potential well in which they are entrapped,
they also must exhibit $|R|>1$. At each reflection (be it of a
particle or antiparticle) pair creation occurs. Again it would be
more correct to say that a non zero probability for pair creation
occurs. We have performed the calculation with the below potential
solution (instead of the antiparticle solution) and a potential
barrier throughout but the physical interpretation is as we have
just described. It is important to note that pair creation occurs
at each reflection and an outgoing wave packet is associated with
each pair production. At least as long as the wave packet width is
much smaller than the barrier(well) width.

We quote below without demonstration the three reflection
scenarios just described. The first line simply lists the step
results of Section II.

{\small
\[
\begin{array}{l|cl|cl|cl}
\mbox{\sc Discontinuity} & \,\,\,& \mbox{\sc Incoming }&\,\,\, &
\mbox{\sc Reflected} &\,\,\, & \mbox{\sc
Transmitted}\\
\mbox{\sc Point} & & \mbox{\sc Wave} & & \mbox{\sc Coefficient} &
& \mbox{\sc Coefficient}\\ \hline \hline &&&&&&\\
 z=0 & & \,\,\,I\rightarrow II  & &(I)\,\,\,\, \dfrac{\alpha+1}{\alpha -1}  & &
(II)\,\,\,\,\,\sqrt{\dfrac{E-m}{E+m}}\, \dfrac{2}{1-\alpha}\\ &&&&&&\\ \hline &&&&&&\\
z=l & & II \rightarrow III & &(II)\,\, \dfrac{\alpha +1}{\alpha
-1}\,\,e^{2iql} & &(III)\,\,\sqrt{\dfrac{E+m}{E-m}}\,
\dfrac{2\,\alpha}{\alpha -1}\,\,e^{i(q-p)l}\\ &&&&&&\\ \hline
&&&&&&\\ z=0 & & \,\,\,I \leftarrow II  & &(II)\,\, \dfrac{\alpha
+1}{\alpha
-1} & & (I)\,\,\,\,\,\,\,\,\sqrt{\dfrac{E+m}{E-m}}\, \dfrac{2\,\alpha}{1-\alpha}\\
&&&&&&\\  \hline
\end{array}
\]
}

\noindent In brackets in columns 3 and 4, we have indicated the
regions in which the reflected and transmitted waves travel.

The build up of the total reflected and transmitted coefficients
is then straightforward. The first few contribution to the
reflection coefficient are ($R_{\1}$ is just the Klein paradox):
\begin{eqnarray*}
R_{\1} & = & \frac{\alpha+1}{\alpha -1}\,\,,\\
R_{\2} & = & \frac{4\,\alpha\,(\alpha+1)}{(\alpha -1)^{^{\3}}}\,e^{2iql}\,\,,\\
R_{\3} & = & \frac{4\,\alpha\,(\alpha+1)^{^{\3}}}{(\alpha
-1)^{^{\5}}}\,e^{4iql}\,\,,\\
 & \vdots  &
\end{eqnarray*}
The subsequent terms are obtained by multiplying by the loop
factor:
\begin{equation}
\label{loopf}
 \left(\frac{\alpha +1}{\alpha -
1}\right)^{\2}\,e^{2iql}
\end{equation}
This loop factor is simply the product of the antiparticle
reflection coefficient at $z=0$ by that at $z=l$ as can be seen
from the table. The increasing phase factor is essential for
yielding the exit times (calculated with the stationary phase
method) of the various wave packets. More about this later in this
Section.

%This phase increment, $\exp[2\,iql]$, predicts a time interval
%$\Delta t$ between successive wave packets, in perfect accord with
%the magnitude of the group velocity of the antiparticle and the
%barrier(well) width, i.e. $\Delta t= 2\,l/q$.

The transmitted coefficients are even simpler since all terms are
related by the loop factor (\ref{loopf}). The first few terms are:
\begin{eqnarray*}
T_{\1} & = & -\,\frac{4\,\alpha}{(\alpha-1)^{^{\2}}}\,e^{i(q-p)l}\,\,,\\
T_{\2} & = & -\,\frac{4\,\alpha\,(\alpha+1)^{^{\2}}}{(\alpha
-1)^{^{\4}}}\,e^{i(3q-p)l}\,\,,\\
T_{\3} & = & -\,\frac{4\,\alpha\,(\alpha+1)^{^{\4}}}{(\alpha
-1)^{^{\6}}}\,e^{i(5q-p)l}\,\,,\\
 & \vdots &
\end{eqnarray*}
The loop factor has modulus greater than one, so that with the
exception of the first reflected wave the probabilities of
subsequent successive wave packets grow geometrically by the
factor
\begin{equation}
\left(\frac{\alpha +1}{\alpha - 1}\right)^{\4}\,\,.
\end{equation}
This growth factor in probability is just the fourth power of the
single step reflection coefficient (which is real and greater than
one). Both of the above series if summed diverge. However, it is
interesting to note that if one does formally sum them, then one
finds exactly the $R_{\B}$ and $T_{\B}$ coefficients of Section
III,
\begin{eqnarray*}
R_{\B} & = & \frac{\alpha+1}{\alpha -1}\,\left\{\,1 +
\frac{4\,\alpha}{(\alpha-1)^{^{\2}}} \,e^{2iql}\,
\sum_{n=0}^{\infty}\, \left[\, \left(\frac{\alpha+1}{\alpha -
1}\right)^{\2}\,e^{2iql}\, \right]^{n}\, \right\}\\
 & = & i\,\frac{\alpha^{\2} -1}{2\,\alpha}\, \sin(ql)\,\,\mbox{\large $/$}\,
 \left[\,\cos(ql) + i\,
\frac{\alpha^{\2}+1}{2\,\alpha}\,\sin(ql)\,\right] \,\,,\\
  T_{\B} & =
& -\,\frac{4\,\alpha}{(\alpha-1)^{^{\2}}}\,e^{i(q-p)l}\,
\sum_{n=0}^{\infty}\, \left[\, \left(\frac{\alpha +1}{\alpha -
1}\right)^{\2}\,e^{2iql}\, \right]^{n}\\\
 & = &
e^{-ip\,l}\,\mbox{\large $/$}\,  \left[\,\cos(ql) + i\,
\frac{\alpha^{\2}+1}{2\,\alpha}\,\sin(ql)\,\right]\,\,.
\end{eqnarray*}
Since the reflected and transmitted series do not converge, we
must conclude that the expressions for $R_{\B}$ and $T_{\B}$ are
{\em not} physical. At most they encrypt the physical
multiwave-packets series. This is in contrast with what happens
for above-barrier diffusion where $|R|$ for the step is less than
one. Then for plane waves (infinite width wave-packets) the
barrier results represent a physical limit and do indeed exhibit
resonance phenomena.

There is a problem with the above results. All exit times except
for the first reflected wave are negative. This is again connected
to time flow under the barrier. Or alternatively because the
antiparticle energy is $-E$ and not $E$. In either case the time
dependent phase factor differs in region II from that in the free
regions. Strictly speaking the stationary plane wave analysis
breaks down because of this. Phase factors due to the time
dependent plane-wave phase must be taken into account.

Fortunately, we do not need to repeat all our analysis because we
know the antiparticle group velocity $\pm q_{\0}/E_{\0}$ and hence
all exit times. These are of course positive with successive times
spaced by $2\,lE_{\0}/q_{\0}$. This is exactly what is obtained if
we take $q\to -q$ in  all the amplitudes. The probability
predictions do not change.

\section*{V.  CONCLUSIONS.}
%%%%%%%%%%%%%%%%%%%%%%%%%%%%%
%%%%%%%%%%%%%%%%%%%%%%%%%%%%%

We have presented in this work the Klein zone analysis for the
step and barrier. In our derivation, we have preferred to use,
 the below-potential "particle" solutions, i.e.  the spinor
 $u^{\c}$. This is not the traditional choice, where the
 $u^{\a}$ spinor is adopted both in the free and potential
 regions\cite{HR81,SMG82,H98,NKM99}. As we have already noteed in
 Section II, this traditional choice has the unpleasant feature of containing,
 in the spinor $u^{\a}$, a denominator which vanishes when
 $E-V_{\0}=-m$ (i.e. at the Klein zone limit where the antiparticle is at rest).
 However, since this vanishing denominator, in
 practice, inverts the small and large
non-relativistic components, the two approaches are compatible and
give the same probabilities. We have also observed that the
barrier region solutions (specifically in their space-time
structure) cannot be the correct antiparticle wave solutions
because they have the wrong group velocity. What is however even
more surprising is the fact that by simple charge conjugation of
these wave functions this problem is not resolved, so even the
charge conjugate solution cannot be the correct antiparticle wave
function.  A similar observation, but based upon different
arguments, has been given by Sakurai\cite{SAK}. Fortunately, this
is not an essential question for our calculations and we have had
no difficulty or confusion when talking about the antiparticles
and their motion, because this is uniquely determined by charge
conservation. However, it is a question which merits further
study.

The Klein paradox is characterized by pair production. As others
have already noted this  can be viewed (a posteriori) as an
anticipation of field theory\cite{KSG04}. The barrier plane wave
solutions seem, at first  sight, not to exhibit  the Klein
paradox. Thus, of concern to us was the fact that there appeared
an inconsistency between the standard barrier result and a two
step calculation. It was also impossible to interpret the  plane
wave barrier results in terms of pair creation. One of our
principle objectives in this work has been to confront these two
approaches. The two step (or more in general multiple step
approach) is not a new idea. In previous applications it has been
lauded as a effective calculational technique\cite{A89,TM91}. We,
on the other hand, have emphasized both here and in a previous
paper\cite{MPLA} its interpretation in terms of multiple wave
packets. In our previous work upon above potential barrier
diffusion, confirmation was also obtained with the help of
numerical calculations.

%For plane waves the barrier result is a consequence of destructive
%interference between the overlapping step waves which results in a
%single reflection and single transmission coefficient. This is a
%notable example of the wave nature of particles. It is by no means
%trivial, since the conservation of particle number (charge) means
%that the antiparticle wave function, which exists only within the
%barrier, must yield through interference zero total probability
%for asymptotic times.

With the Klein paradox, our series solution, each term of which
represents a wave packet, is non convergent. The barrier solution
 represents the formal sum of these series and it is therefore non
physical. We have also observed that the correct amplitudes are
the complex conjugate of our listed results, however this has no
effect upon any predicted probabilities. If the barrier (well)
width is much greater than the wave packet widths then each term
in the series can be studied separately. Each term yields a
separate wave packet and specific exit times. The first reflected
wave ($R_{\1}$) is instantaneously reflected. The first
transmitted amplitude ($T_{\1}[q\to-q]$) exits at time $l
E_{\0}/q_{\0}$. Subsequent reflected wave packets emerge with
intervals of $2\, l E_{\0}/q_{\0}$ and the same for the
transmitted wave packets. If we do no make a distinction between
reflected and transmitted waves, then a wave packet emerges, in
either direction, at intervals of $l E_{\0}/q_{\0}$, the
antiparticle transit time across its potential well.

% We recall that the charge conjugate " positive energy", or
%more precisely "above-potential", solutions correspond to the
%antiparticle state because they see a potential opposite to that
%seen by the particle. However, a by-product of our study convinces
%us, in agreement with Sakurai, that this function $\Psi_{c}$ is
%\emph{not} the correct antiparticle wave function. It has, for
%example, the wrong sign for the group velocity.

%Even neglecting this new paradox it is worth noting that these
%antiparticles, while localized in space, have a \emph{continuous}
%energy spectrum. This is a consequence of the fact that their wave
%functions are not required to be zero at the walls of the
%barrier/well or have exponential damped extensions. For
%exponential solutions we must move out of the Klein energy zone.

The most interesting features of our analysis are:\\
(1) Pair production occurs at zero energy cost.\\
(2) The antiparticles produced via pair production are {\em
permanently} trapped within the potential well (barrier) region.
They lie in a Klein zone of their own.\\
(3) These localized antiparticles have a {\em continuous} energy
spectrum. They are technically not "bound states" because their
existence requires a dynamical process - pair product - at each
reflection. With the Dirac equation, we must distinguish between
static bound state solutions and dynamic localized solutions.
The former are present in the tunnelling zone.\\
(4) The correct amplitudes, with the correct exit times, are those
usually calculated by stationary plane wave analysis with the
substitution $q\to -q$.

The barrier results ($R_{\B}[q\to -q]$ and $T_{\B}[q\to -q]$) are
non physical since they imply the sum of a divergent series.
However, it is to be noted that the continual growth of
antiparticles within the barrier/well region is also a
mathematical abstraction. We expect that as the localized
antiparticle density increases, a corresponding decrease in the
barrier potential height $V_{\0}$ occurs, i.e. an attractive space
charge effect should accrue. Such an effect was confirmed
quantitatively some time ago, within the Thomas Fermi
approximation, in the three dimensional case, by B. M\"uller and
J. Rafelski\cite{MR75}

The Dirac equation has had an incredible success rate when its
predictions are correctly interpreted. Its negative energy
solutions lead to the prediction of antiparticles. Zitterbewegung
is connected to the proven existence of the Darwin term in atomic
physics. The Klein paradox predicts the phenomenon of pair
production. We can therefore hope that {\em dynamic localized
energy spectra}, connected to the geometric growth of periodic
particle emissions will be confirmed experimentally.

\newpage

%%%%%%%%%%%%%%%%%%%%%%%%%%%%%%%%%%%%%%%%%%%%%%%%%%%%%%%%%
%%%%%%%%%%%%%%%%%%%%%%%%%%%%%%%%%%%%%%%%%%%%%%%%%%%%%%%%%
\section*{APPENDIX.}
%%%%%%%%%%%%%%%%%%%%%%%%%%%%%%%%%%%%%%%%%%%%%%%%%%%%%%%%%
%%%%%%%%%%%%%%%%%%%%%%%%%%%%%%%%%%%%%%%%%%%%%%%%%%%%%%%%%

The Dirac equation in presence of  an {\em electrostatic
potential} $A_{\mu}=(A_{\0},\boldsymbol{0})$ reads\cite{ZUB}
\begin{equation}
\left( \,i\,\gamma^{\mu}\partial_{\mu} - e\, \gamma^{\0} A_{\0} -
m\,\right)\Psi(\boldsymbol{r},t) = 0\,\,.
\end{equation}
This equation can be rewritten as follows
\begin{equation}
i\,\partial_{t} \Psi(\boldsymbol{r},t) = (H_{\0} + V_{\0})\,
\Psi(\boldsymbol{r},t)
\end{equation}
where
\[H_{\0}= -\,i \,\gamma^{\0} \boldsymbol{\gamma}\cdot \boldsymbol{\nabla} + m \,
\gamma^{\0}\hspace*{.5cm}\mbox{and}\hspace*{.5cm}
V_{\0}=e\,A_{\0}\,\,.\] Here $e$ denotes the charge of the
particle ($e=-|e|$ for the electron). For a stationary solution
$\Psi(\boldsymbol{r},t) \propto \exp[-\,i\,E\,t]$, we obtain
\begin{equation}
H_{\0} \Psi(\boldsymbol{r},t) = (E - V_{\0})\,
\Psi(\boldsymbol{r},t)\,\,.
\end{equation}
Using the Pauli-Dirac set of gamma matrices
\[ \gamma^{\0}=\left(\begin{array}{rr}\, 1 & 0\\
 0 & \,\,-1\end{array} \right)\hspace*{.5cm}\mbox{and}\hspace*{.5cm}
 \boldsymbol{\gamma}=
 \left(\begin{array}{cr} \,\,\,0 &\,\,\,\boldsymbol{\sigma}
  \\
 - \boldsymbol{\sigma} & 0\,\end{array} \right)\,\,,
\]
the covariant normalized spinorial solutions, $u_{\N}$ are
\begin{equation}
\begin{array}{lclcl}
u_{\N}^{(\1,\2)}(\boldsymbol{q}\,;V_{\0})& =
&\sqrt{(E-V_{\0}+m)}\,
\left( \begin{array}{c}  \chi^{(\1,\2)}\\
\displaystyle{\frac{\boldsymbol{\sigma} \cdot
\boldsymbol{q}}{E-V_{\0}+m}}\,\,\chi^{(\1,\2)}
\end{array} \right) & \,\,\,\,\, & \mbox{for $E - V_{\0} >
m$}\,\,,\\& & & & \\
 u_{\N}^{(\3,\4)}(\boldsymbol{q}\,;V_{\0}) & = & \sqrt{(|E-V_{\0}|+m)}\,\left(
\begin{array}{c} -\, \displaystyle{\frac{\boldsymbol{\sigma} \cdot
\boldsymbol{q}}{|E-V_{\0}|+m}}\,\,\chi^{(\1,\2)} \\ \,\,\,\,\,\,
\chi^{(\1,\2)}
\end{array} \right)& \,\,\,\,\,& \mbox{for $E - V_{\0} < -m$}\,\,,
\end{array}
\end{equation}
where
\[(E-V_{\0})^{^{\2}}
-\,\boldsymbol{q}^{\2} =
m^{\2}\,\,\,\,\,,\,\,\,\,\,\,\,\chi^{(\1)}=\left(
\begin{array}{c} 1
\\0\end{array} \right) \hspace*{.5cm}\mbox{and}\hspace*{.5cm}
\chi^{(\2)}=\left(
\begin{array}{c} 0 \\1\end{array} \right)\,\,.
\]
The two set of solutions $u_{\N}^{(\1,\2)}$ and
$u_{\N}^{(\3,\4)}$, sometimes referred to, imprecisely,  as
"positive" and "negative" solutions, are in fact not determined by
the sign of $E$, but by whether $E>V_{\0}+m$ or $E<V_{\0}-m$.
Hence, $E$ may be fixed but the solutions depend upon whether in
any given region the energy is above or below the potential. For
each of the previous spinors there are in general two separate
solutions corresponding to opposite momentum. In the text of this
paper, we employ the one-dimensional unnormalized spinors $u$,
\begin{equation}
\begin{array}{lclcl}
u^{(\1,\2)}(q\,;V_{\0})& = &\left( \begin{array}{c}  \chi^{(\1,\2)}\\
\,\,\,\,\displaystyle{\frac{q}{E-V_{\0}+m}}\,\,\chi^{(\1,\2)}
\end{array} \right) & \,\,\,\,\, & \mbox{for $E - V_{\0} >
m$}\,\,,\\& & & & \\
 u^{(\3,\4)}(q\,;V_{\0}) & = & \left(
\begin{array}{c} \,\,\,-\, \displaystyle{\frac{q}{|E-V_{\0}|+m}}\,\,\chi^{(\1,\2)} \\
\,\,\,\,\,\, \chi^{(\1,\2)}
\end{array} \right)& \,\,\,\,\,& \mbox{for $E - V_{\0} < -m$}\,\,.
\end{array}
\end{equation}
In free space, $V_{\0}=0$, the above spinors become
\begin{equation}
\begin{array}{lclcl}
u^{(\1,\2)}(p\,;0)& = &\left( \begin{array}{c}  \chi^{(\1,\2)}\\
\,\,\,\,\displaystyle{\frac{p}{E+m}}\,\,\chi^{(\1,\2)}
\end{array} \right) & \,\,\,\,\, & \mbox{for $E >
m$}\,\,,\\& & & & \\
 u^{(\3,\4)}(p\,;0) & = & \left(
\begin{array}{c} \,\,\,-\, \displaystyle{\frac{p}{|E|+m}}\,\,\chi^{(\1,\2)} \\
\,\,\,\,\,\, \chi^{(\1,\2)}
\end{array} \right)& \,\,\,\,\,& \mbox{for $E < -m$}\,\,.
\end{array}
\end{equation}

\end{document}